# Negative charge acquisition by isolated single microdroplets from plasma exposure at atmospheric pressure

Nourhan E Hendawy,[1,2*] Harold N McQuaid,[3,§] Somhairle Mag Uidhir,[3] David Rutherford,[4] Declan Diver,[5] Davide Mariotti,[6] Paul Maguire[5,†]

[1]School of Life Sciences, Sussex University, Brighton, United Kingdom

[2]Biaco Energy Ltd, Hassocks, United Kingdom

[3]School of Engineering, Ulster University, Belfast, N. Ireland

[4]Faculty of Electrical Engineering, Czech Technical University, Technická 2, Prague 6 Czech Republic

[5]School of Physics and Astronomy, University of Glasgow, Glasgow, Scotland

[6]Faculty of Engineering, University of Strathclyde, Glasgow, Scotland

## Abstract

Charged liquid microdroplets have generated significant interest recently due to the observation of chemical reaction rate enhancement by orders of magnitude. Droplet charging by plasma irradiation has been observed along with significantly enhanced reaction rates in liquid. In this paper, we measure the charge on fixed size (47 μm) individual droplets, exposed for approximately 4 ms to a low temperature RF-driven plasma operated at atmospheric pressure. The measured charge values (0.8 – 2.0 x $10^7$ electrons) approached the Rayleigh limit for the given droplet diameter. Results were compared with finite element simulations of plasma charging which provided estimates of surface electric fields and charge fluxes to the droplet surface and helps advance the development of a theoretical framework for plasma-charged microreactors.

## Introduction

Interest in charged microdroplets has recently received considerable attention. Their unique chemical reaction environment offers increasing significance in many scientific and technological fields from materials and drug synthesis to biological medicine [1–6]. Greatly enhanced chemical reaction rates, by many orders of magnitude, which occur in droplets within a specific size range have generated widespread interest for precision chemical synthesis [1,7–9]. Reaction barriers that are extremely challenging in bulk liquids have been overcome, possibly due to high surface electric fields, with important climate-related consequences for high-energy industrial processes such as ammonia production and $CO_2$ transformation [1,3,10–13]. Microdroplets have also been found to spontaneously induce redox reactions with implications for chemical processes including those in living systems, such as photosynthesis and respiration [14–16].

The high surface to volume ratio and the effects of partial solvation on reaction rates are potentially important as well as the suspected presence of surface charge and electric fields which can lead to effects such as ultrathin electronic double layers; ion separation, confinement and alignment; enhanced concentration gradients and diffusive flows; extreme pH conditions, charge transport and spontaneous redox reactions [5,9,17]. Induced ionisation of gas and vapour molecules, droplet fission and plasma formation have also been postulated as a consequence of surface fields. However charge levels and field intensities are unknown with various alternative electrification and charge transfer mechanisms suggested [4,7,8,13,15,17–19].

Microdroplets are a complex physiochemical system with multiple transient parameters that are difficult to control or measure experimentally. Observed effects are critically dependent on size, with an upper limit of approximately 15 μm diameter, and a lower limit of a few microns, determined by lifetime and technical limits. Acquired surface charge is subject to recombination with gas-phase ions during flight. Droplet size and hence internal conditions evolve during flight, due to evaporation, fission and collisions, limiting experimental exposure typically to the sub-millisecond range. Electric fields, concentration gradients, solvent alignment and charge transfer are sensitive to size through their dependence on surface curvature and surface area to volume ratio. Thus, the development of the comprehensive theoretical framework is needed to fully exploit microdroplets capabilities.

*nourhan.hendawy@biaco.energy
§h.mcquaid@ulster.ac.uk
†paul.maguire@glasgow.ac.uk

Microdroplet charging by plasma offers an alternative or complementary approach that could allow greater elucidation of fundamental characteristics as well as enhanced technological capability. On entry into a low-temperature non-equilibrium plasma region, the high electron temperature ensures an initial net flux of high mobility free electrons from the gas phase to the droplet surface, creating a layer of surface charge and an associated electric field [20]. After the droplet floating potential is established, over a few nanoseconds, the electron and positive ion flux are equalized. Thereafter the electrical characteristics remain reasonably constant over a known, and controllable, time period determined by plasma length and droplet velocity, assuming a uniform plasma density and temperature. A suitable plasma is required to operate at atmospheric pressure with low gas temperature to limit droplet evaporation and hence a relatively high gas flow and restricted plasma size is inevitable, presenting a challenge for injection of microscale droplets and charge measurement [21,22]. Recently we reported charge measurements on injected microdroplets into a similar RF-driven plasma as used here [23]. A stream of droplets of varying size and velocity was used and average charge determined. Also, an extremely high rate of in-droplet $H_2O_2$ synthesis was observed, demonstrating the remarkable microreactor capabilities of such droplets. In this paper, we measure the charge on single isolated droplets, of fixed size, in an enclosed flow capillary, using an exterior coaxial ring electrode, for which an electrostatic model was developed. Droplet trajectory, screening and neutralisation were also investigated and the resultant charge values compared to finite element simulation of charging fluxes within the plasma, based on a fully hydrodynamic drift-diffusion model.

## Experimental

Individual microdroplets were generated via a MicroFab (MJ-ATP-01) piezoelectric generator at a rate of 20 Hz, with orifice diameter of 20 µm, leading to an average microdroplet diameter of 47 µm. They were delivered to an RF-driven (13.56 MHz) atmospheric pressure plasma via a 2 mm (ID) quartz capillary tube, interfaced with a gas mixing manifold to allow injection of the plasma gas with limited turbulence. The low temperature plasma was generated between two exterior concentric Cu electrodes, ~2 mm apart and operated in He gas with controlled input gas flows from 0.7 to 1.3 sLm, Figure 1. The acquired charge on the microdroplet, due to plasma exposure, was detected using a thin coaxial ring electrode, located outside the capillary and downstream from the plasma, which was surrounded by a faraday shield to minimize RF induced noise. The current pulse signal was amplified via a low noise charge amplifier (Amptek A250F, A250) and captured on a oscilloscope (Tektronix MDO34).

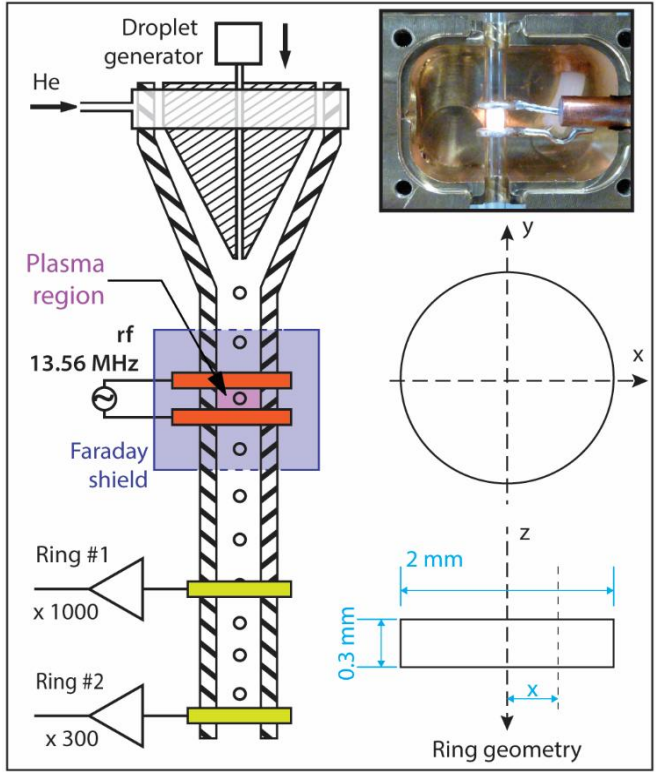

**Figure 1. Experimental setup**

## Theory

The induced current is determined using Ramo – Shockley theorem by assuming the ring electrode of radius a and length w, with w << a, as a series of slices. For a charged droplet travelling through a ring electrode, at an offset x from the radial centre, then the distance, d, between droplet centre at coordinate (x, 0, z) and point ζ on ring, at coordinate (acosθ, asinθ, ζ) is given by

$$d(x,z,\zeta,\theta) = \sqrt{(z-\zeta)^2 + x^2 + a^2 - 2ax\cos\theta} \tag{1}$$

Ignoring the cosθ term allows a simple derivation of closed form expressions for induced charge, $Q_i(t)$, and current, I(t). However, for offset values x > 0.6a, this leads to significant error. Therefore

$$Q_i(t) = -\frac{Q}{2\pi w}\int_{-w/2}^{+w/2}\int_0^{2\pi}\frac{d\theta\, d\zeta}{\sqrt{(vt-\zeta)^2 + x^2 + a^2 - 2ax\cos\theta}} \tag{2}$$

leading to

$$Q_i(t) = -\frac{2Q}{\pi w}\int_{-w/2}^{w/2}\frac{1}{\sqrt{(vt-\zeta)^2+(a+x)^2}}\, K\!\left(\sqrt{\frac{4ax}{(vt-\zeta)^2+(a+x)^2}}\right) d\zeta \tag{3}$$

where K is the complete elliptic integral of 1st kind. From I(t) = $dQ_i/dt$, the closed-form expression for the current pulse is

$$I(t) = \frac{2Qv}{\pi w}\int_{-w/2}^{w/2}\left[\frac{rK(k)}{(r^2+(a+x)^2)^{3/2}} + \frac{2ax\,r}{(r^2+(a+x)^2)^{5/2}}\left(\frac{E(k)}{1-k^2} - K(k)\right)\right]d\zeta,$$

$$r = vt-\zeta, \quad k = \sqrt{\frac{4ax}{r^2+(a+x)^2}} \tag{4}$$

where E(k) is the complete elliptic integral of the 2nd kind. This is solved numerically, in Matlab, using dζ = w/300. Ideally, the pulse amplitude is dependent on one unknown, Q, provided the droplet velocity is known. However, the ring response becomes sensitive to the axial offset, x, once the droplet is sufficiently close to the wall, Figure 2. Furthermore, for droplet trajectories where dx/dz > 0, pulse asymmetry is introduced, Figure 3.

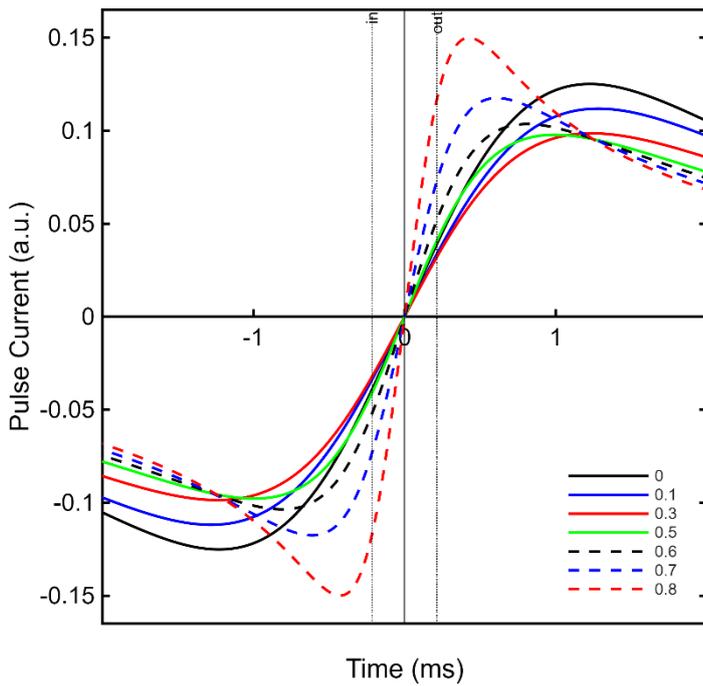

Figure 2: Variation in pulse width and amplitude for various axial offset values, up to x = 0.8a

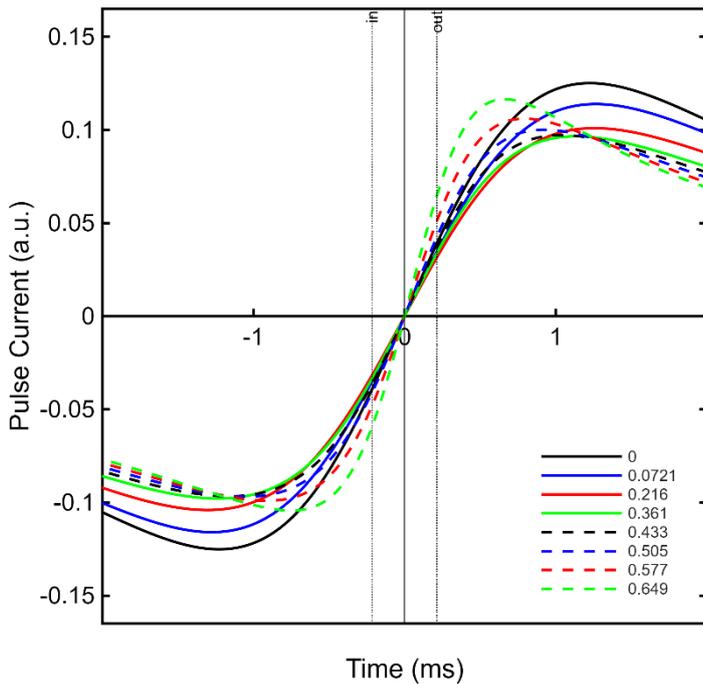

Figure 3: Variation in pulse width and amplitude for various droplet trajectories where dx/dz > 0.

A comparison of pulse height and peak – peak width between a fixed axial trajectory, normal to ring axis, and one that drifts off-normal is given in Figure 4, indicating a close similarity. As the offset value increases, the peak height initially decreases but beyond x/a > 0.5, it increases while the pulse also narrows.

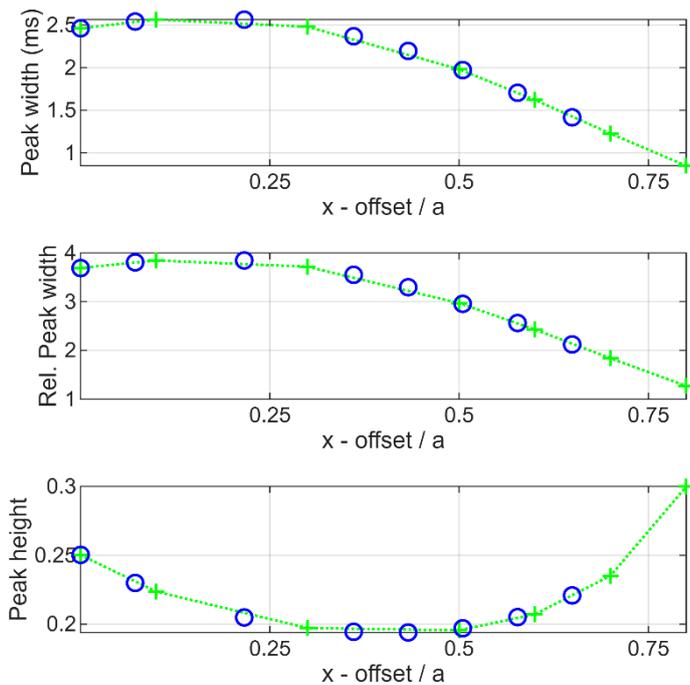

Figure 4: Comparison of pulse parameters, height and peak – peak width, between normal and off-normal trajectories. Relative peak width is with reference to transit time within the ring.

The off-normal trajectory gives rise to an asymmetrical pulse where, for x/a > 0.5 and increasing within the ring, Figure 5, the negative peak amplitude is reduced compared to the positive. However, the total area of the negative pulse is greater than the positive.

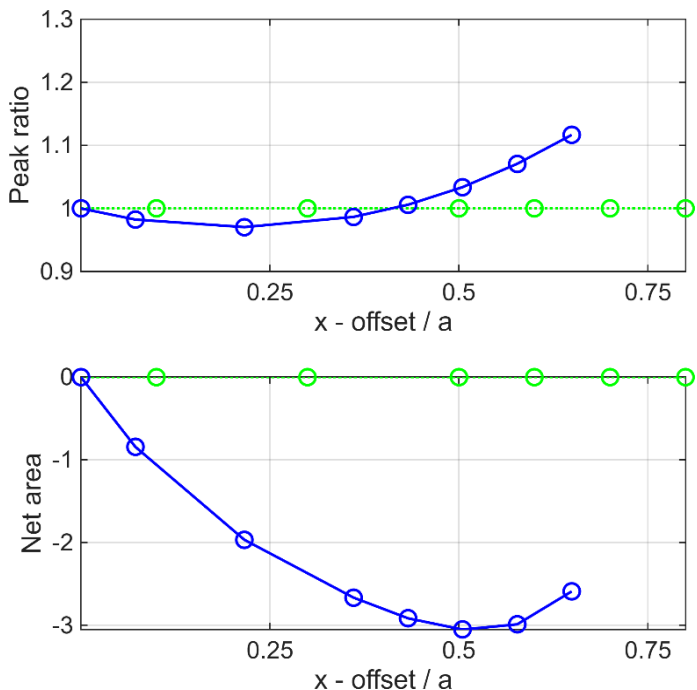

Figure 5: Comparison of pulse parameters, between normal and off-normal trajectories – with x increasing within the ring. The peak ratio represents ratio of positive and negative peak heights while the net area is the integral of the pulse.

The microdroplet velocity is determined from imaging in one plane (y-axis) before and after the plasma region. It is also calculated from the time delay between pulses using a pair of ring electrodes. Simultaneous imaging in the second plane (x-axis) is not possible. Therefore, any x-offset value is estimated by comparing the observed velocity to that expected from the gas velocity, assuming laminar flow conditions and that the droplet velocity has equilibrated to that of the gas. Gas velocity under laminar flow conditions can generally be determined from the upstream set flow values. However, the mixing manifold is designed to minimise back pressure and turbulence while also allowing manual adjustment of the droplet trajectory. As a consequence, capillary gas flow downstream of the manifold is less than the set flow and is therefore measured downstream via liquid displacement tests. At the experimental set flow of 0.7 sLm, the capillary flow value was measured at 0.4 sLm.

The impact of the plasma environment on the ring electrical response also needs to be considered. Beyond the plasma electrodes, a plasma afterglow region is expected where the plasma density decays to zero over a distance $L_{pa}$. A ring electrode situated a distance $L_R$ from the nearest plasma electrode, where $L_R < L_{pa}$, will detect a reduced charge on the droplet due to Debye shielding. The effective charge measured is, to a first approximation, given by

$$Q_{eff} = Q e^{-\frac{s}{\lambda_D}} \qquad (5)$$

where s is the distance between droplet and ring and $\lambda_D$ is the Debye length, given by

$$\lambda_D = \sqrt{\frac{\epsilon_0 k T_e}{n e^2}} \qquad (6)$$

To minimise this effect requires $s/\lambda_D \ll 1$ i.e. the ring is placed beyond the afterglow region. While the precise extent of the afterglow is unclear, an approximation is obtained from the plasma luminance decay profile obtained under various plasma conditions. Current measurements, obtained from a microscale probe in direct contact with the plasma effluent, also provide an indication of the extent of the afterglow region where the current direction changes from negative (towards plasma) to positive. Beyond the afterglow therefore, a region of low-density positive ion space charge is indicated, likely arising as a result of the net loss of more mobile electrons once outside the plasma. This background space charge will also result in screening of the droplet charge. The net charge visible on the ring is that of the droplet minus the total positive charge between droplet and ring. To a first approximation this is given by the term

$$Q_+ = n_+ A s \qquad (7)$$

where A is the area of the capillary cross-section and $n_+$ is the average ion density. This screening effect is expected to decrease on approach and reach a minimum when the droplet is within the ring. Another effect arises in that the presence of a background positive space charge will partially neutralise the droplet charge through ion recombination. Charge reduction due to transit through a space charge region has previously been considered as the swept charge i.e. $n_+ A s$ [24]. However this assumes the relative velocity between particle and space charge is significant, which is unlikely to be the case here. Instead, the random ion flux to the droplet surface is the dominant factor, such that the reduction in charge value is given by

$$\Gamma A = e n_{ion} \pi R^2 \sqrt{\left(\frac{8 k_B T_{ion}}{\pi m_{ion}}\right)} s^{-1} \qquad (8)$$

where $\Gamma$ is the ion flux, A the surface area and R the droplet radius. The magnitude of charge reduction between plasma and ring is estimated using a value of $n_+$ determined from the fit to the current pulse, from (7). The pulse amplitude on a 2nd ring electrode also provides an estimate.

## Results

To establish the optimal position for the ring, as close as possible to the plasma but outside the afterglow region, Figure 6, the length of the afterglow was estimated electrically and from the luminance profile for various plasma power and flow conditions. The luminous length, in the absence of droplets was observed to increase linearly with power, reaching 20 mm at 5W but reduced to 10 mm at 5W, on introduction of droplets, Figure 7. The electrical response indicates a positive ion current flow (from the plasma) at low powers, with the appearance of electron current flow, i.e. a reduction in positive current, occurring above a threshold power value, which varies with distance. Ultimately at high enough powers, the current is negative and indicates the electrode is in full contact with the plasma afterglow. In Figure 8, the effluent current versus power is shown for various plasma to electrode distances, for plasmas without droplets, which confirms the luminance image conclusions. The ring electrode was therefore positioned at a distance of 10 mm from the nearest (ground) plasma electrode. A typical current pulse for a single droplet is shown in Figure 9, with a peak positive voltage of ~150 mV, equivalent to 150 pA. The time interval between negative and positive pulse peaks is ~1 ms.

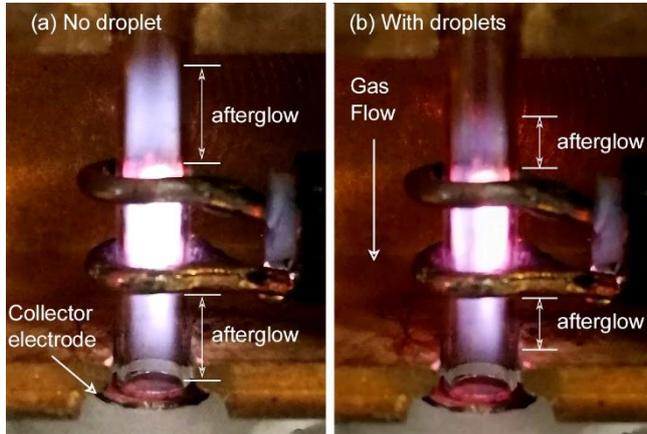

Figure 6: Plasma afterglow - luminous length extends beyond the plasma electrodes

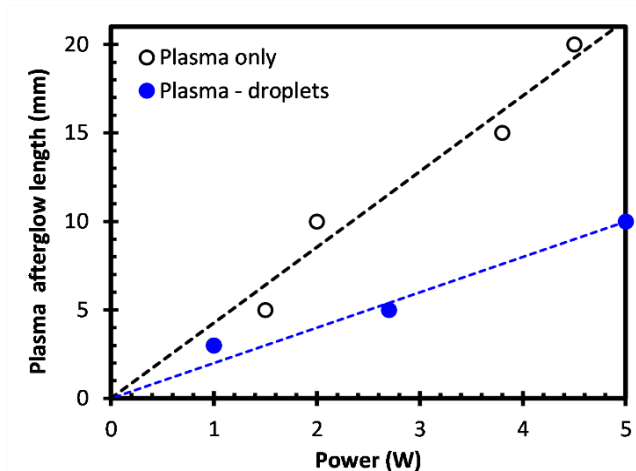

Figure 7: Plasma afterglow length (luminous) versus absorbed power for plasma with and without droplets

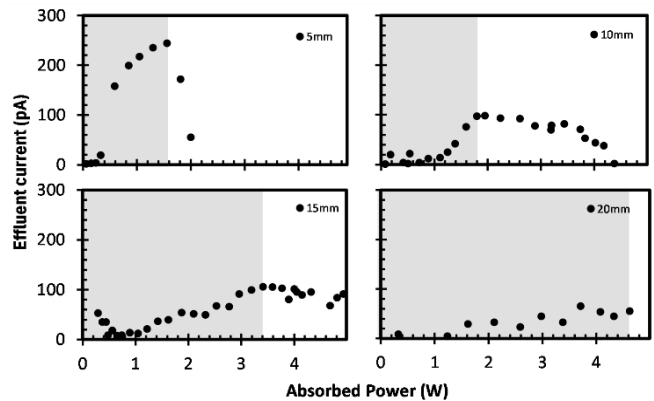

Figure 8: Effluent current versus power indicating the variation in power threshold for coupling to plasma afterglow for various distances between plasma electrode and current measurement electrode. The shaded region represents the safe power range to ensure the electrode is outside the afterglow region.

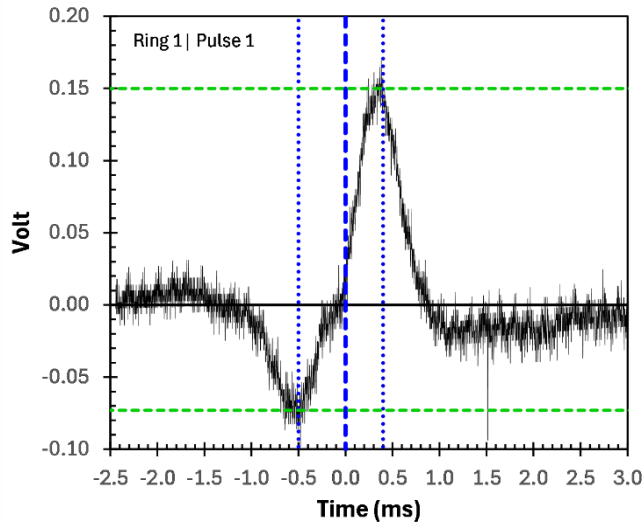

Figure 9: Typical ring voltage pulse due to passage of an individual charged microdroplet through the ring. The signal mean has been zeroed to remove a negative DC offset. The pulse width (peak – peak) is ~1 ms. Scaling factor: 1 V = $10^{-9}$ A. The microdroplet/pulse frequency was 20 Hz. For analysis, signal averaging is carried out using 200 pulse signal acquisitions.

Droplet size and velocity was determined from image measurements at the piezo tip and in the plasma region (without plasma ignited) for various generating and plasma conditions, Figure 10. The observed droplet diameter was 47 μm ± 5% at release from the piezo tip and remained constant beyond the ring, within the imaging resolution. For a plasma evaporation rate constant, $c_{evap}$, of $3 \times 10^{-8}$ m$^2$ s$^{-1}$, previously determined for a similar plasma system, and assuming $D_{final}^2 = D_0^2 c_{evap} t$, the expected reduction is < 2 μm [22]. For a 20 Hz droplet stream, the observed average velocity was 0.45 ± 3% m s$^{-1}$ in a measured gas flow of 0.4 sLm. In this flow scenario, the Reynolds number is small (< 1), Stokes drag dominates and the droplet terminal velocity (~ $v_{gas}$) is expected to be reached within 3 mm of the He gas inlet, which is far upstream (~20 mm) from the plasma and ring. Assuming laminar flow conditions at the ring, i.e. a parabolic velocity profile, the peak velocity at zero offset (x = 0) is 4.2 ms$^{-1}$, while for an offset x = 0.67a, the velocity is 0.45 ms$^{-1}$, equal to the measured droplet velocity. A droplet velocity of 0.44 m s$^{-1}$ was also estimated from the time interval between pulses on a second ring located a fixed distance (21 mm) away.

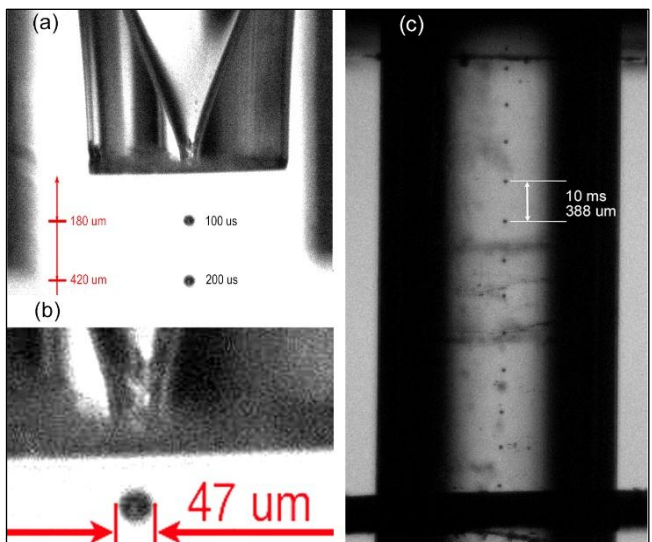

Figure 10: Droplet size and velocity determination. (a) droplets exiting piezo generator tip at a velocity of 2.1 m s$^{-1}$ into still air, (b) enlarged image at tip, droplet diameter of 47 μm. Dimensions are calibrated against vernier calliper measurements of the tip enclosure (806 μm), (c) droplets (100 Hz) within plasma region, with He flow but without plasma ignited showing 388 μm separation, for a set interval of 10 ms.

Simulated current pulses were generated for a droplet velocity of 0.45 m s$^{-1}$ and an x-offset value of 0.68a, as observed experimentally. Also, since the pulse is asymmetric, this indicates a variation in x-offset during flight. In Figure 11 the comparison between simulated and measured currents is shown. Peak heights are matched using charge values 9 x 10⁶ electrons (fixed x-offset) to 2 x 10⁷ electrons (variable offset).

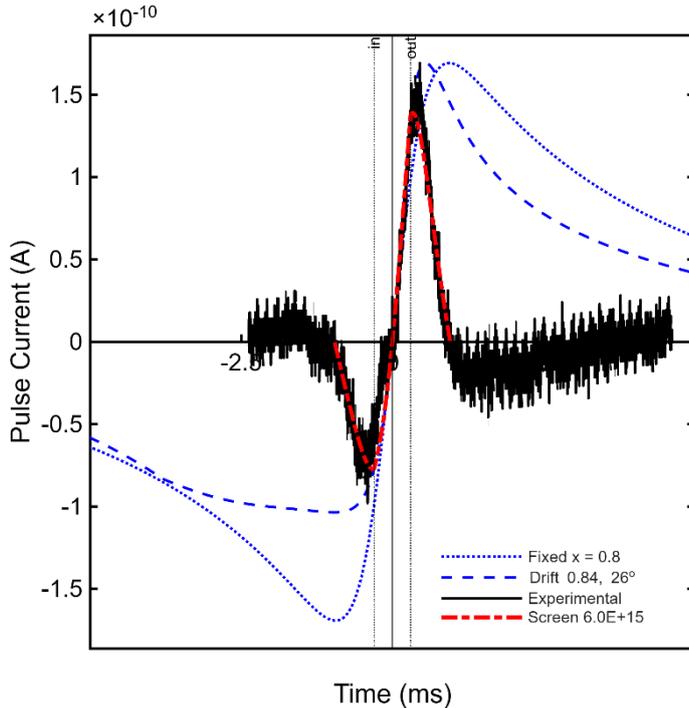

**Figure 11: Comparison between measured current and simulation for fixed and variable x-offsets and for positive ion charge screening with an ion density of 5 x 10¹⁵ m⁻³.**

The rapid decay in pulse amplitude may be attributed to screening by the background positive ion density, which was adjusted to 6 x 10¹⁵ m⁻³ to provide an optimal fit to the measured pulse shape. The effect of positive ion density on droplet charge neutralisation was also investigated by comparing pulse amplitudes on two rings, separated by a gap of 21 mm. Over this distance, the observed decay (72%) indicates, from (8), an ion density of 1.8 x 10¹⁴ m⁻³ and the expected decay between plasma and ring is ~17%. A similar calculation was carried out for pulses obtained from a high-rate droplet stream (50 kHz) with a single ring electrode positioned at various distances, up to 60 mm, from the capillary output. In this case, the droplet diameters followed a lognormal distribution (mean 13 μm) with velocities in the range of 3 - 14 m s⁻¹. Figure 12 shows an example fit to the experimentally observed charge decay for a droplet velocity of 6 m s⁻¹ and an ion density of 2 x 10¹⁵ m⁻³.

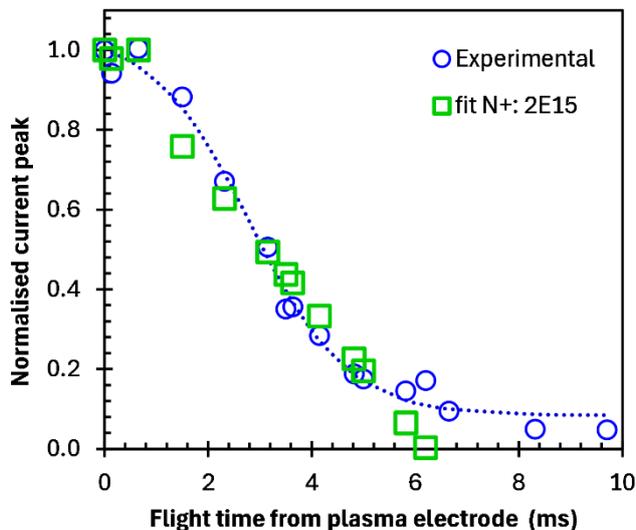

**Figure 12: Relative current pulse amplitudes versus time of flight from plasma electrode. Droplet velocity 6 ms⁻¹.**

## Discussion

The estimated value of charge on fixed size individual droplets varied from 0.8 to 2.0 x 10⁷

electrons, depending on the x-offset conditions chosen to match the observed droplet velocity to that of the gas. Due to charge neutralisation by the weak background positive space charge between end of plasma and the ring electrode, measured values are estimated to be ~17% lower that when droplets are in the plasma region. This is a very high level of charge compared to our measurements made previously using a solid plate electrode downstream of the capillary, reaching the Rayleigh limit, i.e. the maximum charge possible, for the droplet size [23]. The two sets of measurements are not expected to be comparable due to variation in plasma conditions for different gas flows, the distributions of charge and velocity values in the latter along with possible higher neutralisation and flow effects around a solid measurement electrode situated in air. Previously, finite element simulations were carried out to estimate the acquired charge, assuming a fully collisional plasma and using electron and positive ion drift – diffusion equations to the droplet surface. These were coupled via Poisson's equation to the floating potential of the charged droplet which was iterated until electron and ion flux were equal. The main parameters were $n_e$, $T_e$, ion mobility, $\mu_{ion}$ and R. For both ring and plate measurement sets, simulation predicted much lower levels of charge than observed. In the latter case, reduction of the ion mobility to account for large slow positive cluster ions offered a possible solution. Such ions are expected due to the layer of saturated water vapour surrounding an evaporating droplet travelling at the same speed as the gas and hence are also likely to be present in the ring measurements. Multiple collisions between $He_2^+$ ions and the evaporated water molecules surrounding the droplet lead to the growth of large positive, $H^+(H_2O)_n$, and negative $OH^-(H_2O)_n$, water clusters ions, the cluster size increasing with vapour content [25–27]. Large positive cluster ions with n up to 55 (~1000 amu) have been reported for a He plasma jet into air [28]. For estimated average values of $n_e = 10^{20}$ $m^{-3}$ and $T_e = 2$ eV, the predicted charge from simulation for R = 24 µm, was 7 x $10^5$ electrons and 1.2 x $10^6$ electrons for mobility values of 2.5 x $10^{-3}$ V $m^{-2}$ $s^{-1}$ ($He_2^+$) and 1.0 x $10^{-5}$ V $m^{-2}$ $s^{-1}$ $(H_2O)_n^+$ respectively. However, a further reduction in mobility is not realistic.

From a technological perspective, charged droplets may have important application, in for example downstream spatial printing of advanced materials. In the case of the higher velocity droplet stream, the neutralisation rate flattens beyond a time-of-flight post plasma of ~8 ms, leaving the droplet charge value at ~$10^6$ electrons, considerably higher than can be achieved by other techniques, e.g. corona or triboelectric charging. In corona discharge systems at atmospheric pressure, ions are primarily responsible for particle charging because high mobility electrons are quickly lost through attachment and recombination in dense gases. Reported values for corona charging range from approximately 100 electrons for 1 µm particles to about 2000 electrons for 7 µm particles [29]. In dusty plasmas, charging measurements are largely limited to nanoparticles at high pressure and microparticles at low pressure. Simulations suggest nanoparticle charges of around 200 electrons for 125 nm diameter particles [30], while low-pressure plasma experiments show values ranging from 10 to $10^4$ electrons for particles up to ~10 µm diameter [29,31–33].

However, the more significant technological potential may be in the exploitation of plasma charged droplets as chemical microreactors. We have previously shown enhancement of reaction rates in droplets by many orders of magnitude compared to bulk techniques for Au nanoparticle formation from metal salt reduction and for $H_2O_2$ synthesis [23,34]. Solvated electron reactions drive the former and are likely an important element of the latter. Therefore, important parameters for microreactors are the electron charge magnitude, charge flux to the surface and plasma time-of-flight. The observed values of charge can be matched in simulation by increasing $n_e$ and $T_e$, typically charge increases by factors of x3 to x5 per decade increase in electron density and similarly for increases in $T_e$ up to 13 eV. To date we have used values of $n_e$ obtained from impedance measurements which assume a uniform density cylindrical plasma. Although spatial or temporal models of plasmas at atmospheric pressure, particularly RF-driven, are absent, we might speculate that spatiotemporal variation in $n_e$ and $T_e$ could be significant, especially near the driven electrode, at the sheath edge and where the droplet may come close, given the x-offset is ~0.7a. In that eventuality, the question is whether the elevated charge acquired in this region remains until the droplet exits the plasma or rapidly equilibrates to the local density and temperature. To evaluate this, we consider a droplet charged near the upper electrode (RF-driven) with local excess $n_e$ and $T_e$ which then passes through the remainder of the plasma region (2 mm, 4 ms) of uniform $n_e$ and $T_e$, as used in simulations. The charge flux density at the droplet surface is constant at a value of

$10^{20}$ m$^{-2}$ s$^{-1}$ to 3 x $10^{22}$ m$^{-2}$ s$^{-1}$ for low and high mobility values. Assuming, for simplicity that the electron flux ~0 until the charge is equilibrated, then sufficient ion neutralisation would occur within only 15 µm of travel. Alternatively, if $n_e$ and $T_e$ are spatially uniform but vary with RF amplitude, the required neutralisation time is >100 times the RF period and hence enhanced plasma charging as a result of temporal spikes in plasma parameters is a plausible possibility.

Finally, chemical reaction rate enhancement in charged droplets, via techniques other than plasma exposure, have indicated a strong size and surface area effect, with surface species concentration and electric fields thought to be implicated. We observe from simulation an increase in charge with diameter which follows a 2nd order polynomial fit. However, the flux density decreases with diameter and hence we can expect that chemical reaction rate enhancement, which likely depends on species concentrations in a shallow surface layer, will also be favoured as the diameter is reduced. Furthermore, the simulated surface electric field also shows a rapid increase for smaller droplets. From a technological perspective, the minimum droplet size will depend on evaporation rate which will determine the lifetime of the reactor [20]. With the RF-driven plasma system we have previously measured evaporation rates approximately ten times higher than that in flowing gas at room temperature. Under these conditions, droplets with initial diameters under 5 µm are totally evaporated after 100 µs plasma exposure time. Using the high-rate droplet source where the large reaction rate enhancements were observed, the mean droplet diameter was 13 µm and while the current single droplet system can produce droplets as small as 15 µm, it was found that most droplets smaller than ~40 µm were lost to the walls at the gas mixer inlet. Future developments will address this issue.

## Conclusions

We have demonstrated single droplets with net negative acquired charges after short exposure to a low temperature RF-driven atmospheric pressure plasma. The measured acquired charge was between 9 x $10^6$ electrons and 2 x $10^7$ electrons which is close to the Rayleigh limit for a 47 µm diameter water droplet. These values are much higher than obtained from simulation and possible reasons include reduced positive ion mobility due to the creation of large heavy water cluster ions in the evaporation halo around the droplet and large temporal enhancement of electron density and temperature with RF amplitude. Overall, the droplet in plasma system is a complex multiphase environment but one that offers exciting opportunities for technological exploitation as chemical microreactors and advanced materials manufacture.

## Acknowledgements

This work was supported by Engineering and Physical Sciences Research Council (Project Nos. EP/K006088/1, EP/K006142/1, EP/K022237/1, EP/R008841/1, EP/T016000/1) and EU COST Actions PlAgri (CA19110) and PlasTHER (CA20114).